\documentclass[a4paper,twocolumn,11pt,accepted=2022-05-17]{quantumarticle}
\pdfoutput=1
\usepackage[utf8]{inputenc}
\usepackage[english]{babel}
\usepackage[T1]{fontenc}
\usepackage{amsmath}
\usepackage{hyperref}
\usepackage{bm}
\usepackage[numbers,sort&compress]{natbib}

\usepackage{graphicx}
\usepackage{amssymb,amsmath}
\usepackage{bm}
\usepackage{dcolumn}
\usepackage{subfigure}
\usepackage{float}
\usepackage[OT1]{fontenc} 
\usepackage{url}
\usepackage{slashed}
\usepackage{color}
\usepackage{verbatim}
\usepackage{txfonts}
\usepackage{soul}
\usepackage{subfigure}
\usepackage{comment}
\usepackage{amssymb}

\usepackage{tikz}
\usepackage{lipsum}

\def\bea{\begin{eqnarray}}
\def\eea{\end{eqnarray}}

\def\ba{\begin{array}}
\def\ea{\end{array}}

\usepackage{environ}

\NewEnviron{myequation}{%
    \begin{equation}
    \scalebox{0.87}{$\displaystyle{\BODY}$}
    \end{equation}
    }

\begin{document}

	\title{Universal Entanglement Transitions of Free Fermions with Long-range Non-unitary Dynamics}
	
	\author{Pengfei Zhang}
	\affiliation{Institute for Quantum Information and Matter and Walter Burke Institute for Theoretical Physics, California Institute of Technology, Pasadena, CA 91125, USA}
	
	\author{Chunxiao Liu}
	\email{chunxiaoliu@ucsb.edu}
	\affiliation{Department of Physics, University of California Santa Barbara, Santa Barbara, CA 93106, USA}
	
	\author{Shao-Kai Jian}
	\email{skjian@brandeis.edu}
	\affiliation{Condensed Matter Theory Center and Joint Quantum Institute,
Department of Physics, University of Maryland, College Park, MD 20742, USA}

	\author{Xiao Chen}
	\email{chenaad@bc.edu}
\affiliation{Department of Physics, Boston College, Chestnut Hill, MA 02467, USA}

\maketitle

\begin{abstract}
		 Non-unitary evolution can give rise to novel steady states classified by their entanglement properties. In this work, we aim to understand the effect of long-range hopping that decays with $\bm{r^{-\alpha}}$ in non-Hermitian free-fermion systems. We first study two solvable Brownian models with long-range non-unitary dynamics: a large-$N$ SYK$_2$ chain and a single-flavor fermion chain, and we show that they share the same phase diagram. When $\bm{\alpha>0.5}$, we observe two critical phases with subvolume entanglement scaling: (i) $\bm{\alpha>1.5}$, a logarithmic phase with dynamical exponent $\bm{z=1}$ and logarithmic subsystem entanglement, and (ii) $\bm{0.5<\alpha<1.5}$, a fractal phase with $\bm{z=\frac{2\alpha-1}{2}}$ and subsystem entanglement $\bm{S_A\propto L_A^{1-z}}$, where $\bm{L_A}$ is the length of the subsystem $\bm{A}$. These two phases cannot be distinguished by the purification dynamics, in which the entropy always decays as $\bm{L/T}$. We then confirm that the results are also valid for the static SYK$\bm{_2}$ chain, indicating the phase diagram is universal for general free-fermion systems. We also discuss phase diagrams in higher dimensions and the implication in measurement-induced phase transitions.
\end{abstract}

\section{Introduction}

Unveiling new phases and novel dynamics of quantum many-body systems is one of the most important subjects in condensed matter physics. Recent developments find that a new paradigm exists when we consider non-unitary evolutions. It is found that for a ``hybrid'' quantum dynamics composed of both unitary evolution and projective measurement, if we follow the quantum trajectories, the steady state exhibits a transition between a volume-law entangled phase and an area-law entangled phase by varying the measurement strength \cite{Cao_Tilloy_2019,Li_2018,Li_2019,Skinner_2019,Chan_2019,Bao_2020,Choi_2020,gullans2019dynamical,gullans2019scalable,jian2019measurementinduced,zabalo2019critical,Tang_Zhu_2020,Szyniszewski_2019,Zhang_2020,goto2020measurementinduced,jian2021yang,buchhold2021effective,bao2021symmetry}. Later studies show the entanglement transitions also appear in more generalized non-unitary random evolutions \cite{sang2020measurement,Lavasani_2021,ippoliti2021fractal,lu2021entanglement,jian2020criticality,Ippoliti_2021}. In particular, in free-fermion systems, under non-unitary random evolution, there is a stable critical phase, in which the steady state has power-law correlation functions and logarithmic entanglement entropy, with possible entanglement transitions into area-law phases \cite{bao2021symmetry,alberton2020trajectory,Chen_2020,liu2020non,part_one,part_two,Nahum_2020,Tang_2021,biella2021many,PhysRevB.103.224210,turkeshi2021entanglement,turkeshi2022entanglement}. This criticality is attributed to the existence of Goldstone modes from the spontaneous breaking of the continuous symmetry in the enlarged replicated Hilbert space \cite{bao2021symmetry,part_one,part_two}. 

 Most of these studies focus on quantum systems with local interactions. However, most of the state-of-the-art experimental platforms to simulate the quantum many-body dynamics contains intrinsic long-range interactions. For example, the ultracold atoms in the optical lattices interact with a Van der Waals potential $\sim 1/r^6$ and the dipole-dipole interaction in the NMR system decays even slower $\sim 1/r^3$. These long-range interactions can significantly change the ground state property \cite{viyuela2016topological,viyuela2018chiral} and the quantum dynamics \cite{Hastings_2006,matsuta2017improving,Chen_2019,chen2019finite,zhou2020operator,tran2021liebrobinson,Kuwahara_2020} under unitary evolution. For example, the Lieb-Robinson bound in local spin chains can receive non-trivial corrections, giving rise to rich lightcone structures depending on the strength of the interaction and the local Hilbert space dimension \cite{Hastings_2006,matsuta2017improving,Chen_2019,chen2019finite,zhou2020operator,tran2021liebrobinson,Kuwahara_2020}. Consequently, it 
 is not only of great interest but also urgently necessary to correctly incorporate the effect of long-range couplings in 
 non-unitary evolution. Although several papers of numerical simulations of long-range interacting \cite{block2021measurement,minato2021fate} or non-interacting \cite{minato2021fate} systems under non-unitary evolutions appeared recently, a definite answer of the entanglement properties of { non-unitary long-range-coupled random free fermion dynamics} is still lacking.
 
	\begin{figure}[tb]
		\centering
		\includegraphics[width=1\linewidth]{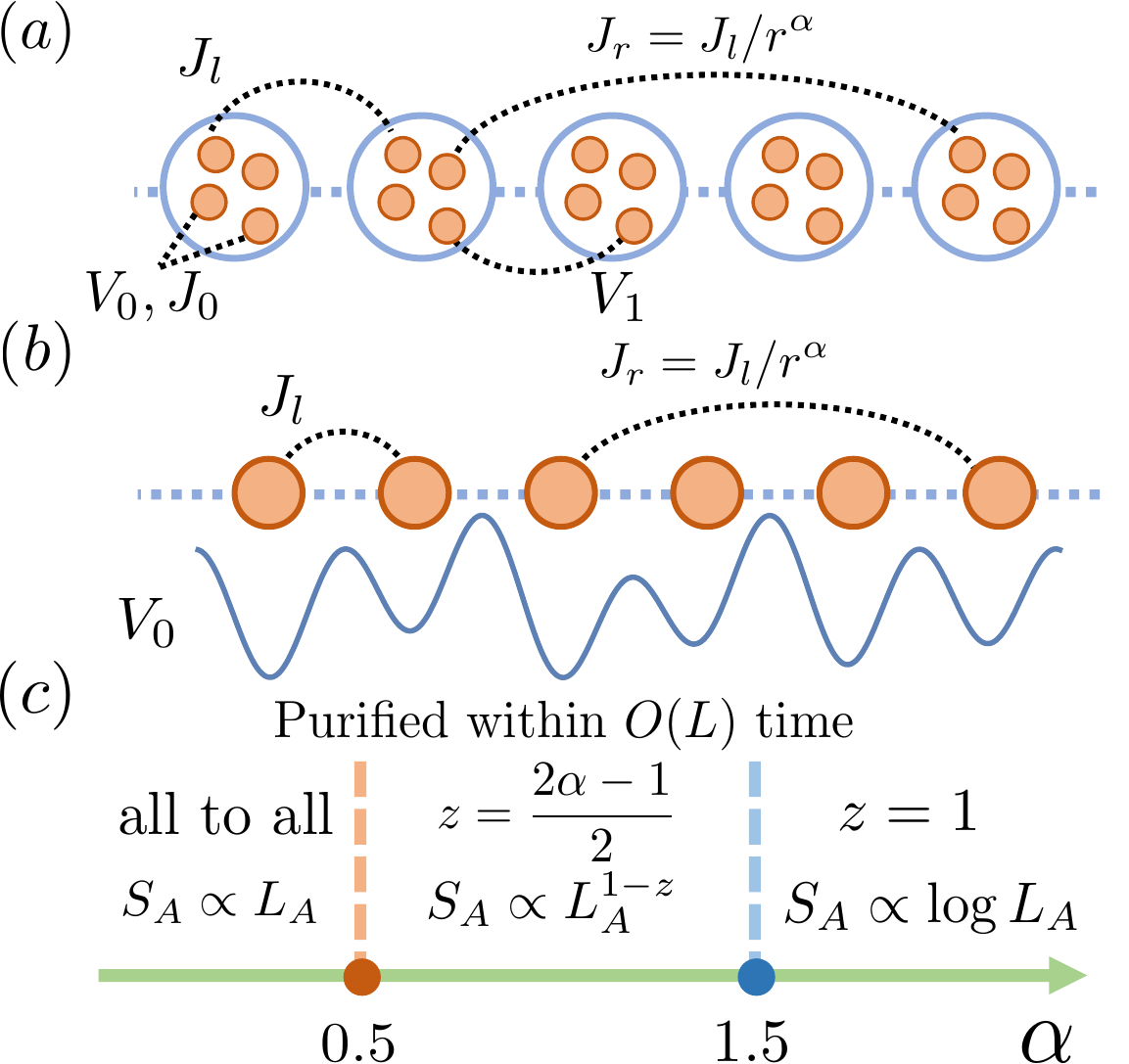}
		\caption{Schematics of (a). The long-range non-hermitian SYK$_2$ chain. The Hamiltonian is either Brownian or static. (b). Single-flavor chain with imaginary random on-site potential. The randomness is time dependent. In both cases, we use $J$ or $V$ for terms in the $H_R$ or $H_I$ respectively. In particular, $J_r$ represents the long-range hopping. (c). The phase diagram is valid in both models, regardless of the local Hilbert space dimension. Here $z$ is the dynamical exponent and $S_A$ is the entanglement entropy of a subsystem with length $L_A$.}
		\label{fig:schemticas}
	\end{figure}

 In this work, we present a detailed study on non-unitary free-fermion systems with long-range hopping $J_r\propto 1/r^\alpha$, where $r$ labels the distance between two sites. We investigate this problem by constructing two { analytical} solvable models in $1$-D: the quadratic Brownian Sachdev-Ye-Kitaev (SYK) chain \cite{kitaev2015simple,maldacena2016remarks,Sachdev_Ye,gu2017local,davison2017thermoelectric,chen2017competition,song2017strongly,zhang2017dispersive,jian2017model,chen2017tunable,saad2018semiclassical,sunderhauf2019quantum} with $N\rightarrow \infty$ fermions per site, and a single-flavor $N=1$ Brownian chain. { In both models, the non-unitary evolution can be treated as the free fermion dynamics subject to continuous weak measurements. We will show that such non-unitary dynamics with non-local hopping can stabilize phases with non-trivial correlation and entanglement structure.}
 
{ The SYK$_q$ model describes $N$ Majorana modes with random $q$-body couplings~\cite{kitaev2015simple,maldacena2016remarks,Sachdev_Ye}, which is solvable under the large-$N$ expansion. The SYK model has been simulated using nuclear spin chains \cite{luo2019quantum}, and a solid-state realization has been proposed using a graphene flake \cite{chen2018quantum}. Later, different generalizations of the SYK model have been studied, including non-Hermitian Brownian SYK chains in which measurement induced phase transitions can be analyzed using the effective action approach \cite{liu2020non,part_one,part_two}. Motivated by these developments, here we use Brownian SYK chains to understand entanglement properties of the steady state under long-range non-unitary evolutions. To test whether the result is sensitive to the large-$N$ limit, we further study the single-flavor Brownian chain using nonlinear master equations \cite{chen2020emergent}. We find that all results match in these models, indicating that the physics is independent of the local Hilbert space dimension.} For $\alpha>0.5$, they show critical behavior and can be further separated into two phases (see FIG. \ref{fig:schemticas}). For $\alpha \!>\! 1.5$, the long-range hopping decays rapidly enough and the result is the same as the short-range hopping case with dynamical exponent $z=1$ and logarithmic entanglement entropy. For $0.5 \!<\!\alpha \!<\! 1.5$, the system is in a fractal phase \cite{ippoliti2021fractal} with subsystem entanglement $S_A\propto L_A^{1-z}$, where $L_A$ is the length of the subsystem $A$ and $z=(2\alpha-1)/2$. We further confirm that the same result holds for static SYK$_2$ chain and we propose the phase diagram to be universal for generic non-unitary random free-fermion systems. We also analyze the purification dynamics, where we find the entropy of the system decays as $L/T$ for any arbitrary $\alpha$.

\section{Large-\texorpdfstring{$N$}{TEXT} Model}
 We first consider the long-range non-Hermitian SYK$_2$ chain with $N$ Majorana fermions $\chi^i$ on each site (see FIG. \ref{fig:schemticas}). The Hamiltonian $H=H_R-iH_I$ reads
\begin{myequation}\label{eqn_HSYK}
\begin{aligned}
H_R(t)&=\sum_{x,ij}i\tilde{J}_{ij}^{x}(t) \chi^i_x\chi^j_x/2+\sum_{r\geq 1}\sum_{x,ij}i\frac{J_{ij}^{x,x+r}(t)}{r^\alpha} \chi^i_x\chi^j_{x+r},\\
H_I(t)&=\sum_{x,ij}\left[iV_{ij}^{x,x+1}(t) \chi^i_x\chi^j_{x+1}+i\tilde{V}_{ij}^{x}(t) \chi^i_x\chi^j_x/2\right].
\end{aligned}
\end{myequation}
Here $i,j=1,2...N$ labels the Majorana modes $\chi$ on each site and $r$ labels the hopping distance. Here $H_I$ contains intra-site and nearest neighbor hopping, and $H_R$ contains intra-site coupling and long-range hopping. $\tilde{J}_{ij}^{x}$, $J_{ij}^{x,x+r}$, $V_{ij}^{x,x+1}$ and $\tilde{V}_{ij}^{x}$ are independent random Gaussian variables with zero expectation value. To be concrete, we focus on the Brownian case with variance
\begin{myequation}
\begin{aligned}
&\overline{\tilde{J}_{ij}^{x}(t)\tilde{J}_{ij}^{x}(0)}=\frac{J_0\delta(t)}{N},\ \ \ \ \overline{V_{ij}^{x,x+1}(t)V_{ij}^{x,x+1}(0)}=\frac{V_1\delta(t)}{2N},\\&\overline{\tilde{V}_{ij}^{x}(t)\tilde{V}_{ij}^{x}(0)}=\frac{V_0\delta(t)}{N},\ \ \ \ \overline{J_{ij}^{x,x+r}(t)J_{ij}^{x,x+r}(0)}=\frac{J_l\delta(t)}{2N}.
\end{aligned}
\end{myequation}
For simplicity, we set $J_l=J_0$.  Physically, quantum trajectories undergoing non-unitary evolution can be realized by continuous forced measurements as discussed in Appendix \ref{app:exp}. There are also other related studies of measurements and feedback in quantum many-body systems \cite{ashida2016quantum,ashida2017parity,mazzucchi2016quantum,mazzucchi2016quantum2,dhar2016measurement,ivanov2020feedback,buonaiuto2021dynamical}. In particular, in \cite{ivanov2020feedback}, authors find feedback, which gives rise to effective long-memory couplings, can change the critical behaviors.

We are interested in analyzing the steady state under non-unitary evolutions. The system is prepared in some initial state $|\psi_0\rangle$. At time $t$, the wave function evolves as $|\psi(t)\rangle=e^{-iHt}|\psi_0\rangle/\sqrt{\langle\psi_0|e^{iH^\dagger t}e^{-iHt}|\psi_0\rangle}$. Here we keep any time-ordering operator implicit. {Recent studies show phases can be classified according to their entanglement properties~\cite{Cao_Tilloy_2019,Li_2018,Li_2019,Skinner_2019,Chan_2019,Bao_2020,Choi_2020,gullans2019dynamical,gullans2019scalable,jian2019measurementinduced,zabalo2019critical,Tang_Zhu_2020,Szyniszewski_2019,Zhang_2020,goto2020measurementinduced,jian2021yang,buchhold2021effective,bao2021symmetry,sang2020measurement,Lavasani_2021,ippoliti2021fractal,lu2021entanglement,jian2020criticality,Ippoliti_2021,alberton2020trajectory}. }We consider the second R\'enyi entanglement entropy, the definition of which includes two replicas of the original system. We choose a subsystem $A$ with length $L_A$. The purity of the subsystem $A$ reads 
\begin{myequation}\label{eqn_purity}
P_A=\text{tr}_A(\rho_A)^2=\frac{\text{tr}_A\left(\text{tr}_B~e^{-iHt}\left|\psi_0\right>\left<\psi_0\right|e^{iH^\dagger t}\right)^2}{\left(\left<\psi_0\right|e^{iH^\dagger t}e^{-iHt}\left|\psi_0\right>\right)^2},
\end{myequation}
where $\text{tr}_{A/B}$ denotes the partial trace of subsystem $A$ or $B$, and the second R\'enyi entropy can be obtained as $S^{(2)}_A(t)=-\overline{\log P_A(t)}$. Previous studies \cite{part_one,part_two} show Keldysh squared correlators, which probe the correlation between two replicas, can also distinguish different phases. It is defined as $F=\sum_{ij}\overline{\left<\chi_x^i\chi_0^j\right>^2}/N$, where
\begin{myequation}
\left<\chi_x^i\chi_0^j\right>\equiv \underset{t\rightarrow \infty}{\text{lim}}\frac{\left<\psi_0\right|e^{iH^\dagger t} e^{-iHt/2} \chi_x^i \chi_0^je^{-iHt/2}\left|\psi_0\right>}{\left<\psi_0\right| e^{iH^\dagger t} e^{-iHt}\left|\psi_0\right>}.
\end{myequation}

Using the fact that the saddle-point solution in the large$-N$ limit is the disorder replica diagonal in the SYK-like models \cite{maldacena2016remarks,kitaev2018soft,gu2020notes}, both $F$ \cite{part_one} and $S^{(2)}_A$ \cite{liu2018quantum,gu2017spread,huang2019eigenstate,10.21468/SciPostPhys.8.6.094,haldar2020Renyi,zhang2020entanglementy,chen2020Replica,liu2020non,shao2021note,part_two} can be expressed as a path-integral with two replicas. We begin with the evaluation of $\overline{(\left<\psi_0\right| e^{iH^\dagger t} e^{-iHt}\left|\psi_0\right>)^2}$. The contour contains four branches (labeled by $1$-$4$), with two forward evolutions ($1$, $3$) and two backward evolutions ($2$, $4$). Here $1$ and $2$ ($3$ and $4$) belong to the same replica. {The standard derivation \cite{maldacena2016remarks} leads to the $G$-$\Sigma$ action:
\begin{equation} \label{sm:full_action_main}
\begin{aligned}
\frac{I}N =& -\sum_x \frac12 \text{tr} \log \Big( (-1)^{a+1} \delta^{ab} \partial_t - \Sigma_x^{ab} \Big) \\& + \int_{t_1,t_2} \Big[ \frac12 \Sigma_x^{ab} G_x^{ab} +   \frac{(-1)^{a + b}\delta(t_{12})}4 \\&\times[J_0 (G_x^{ab})^2 
+\sum_r\frac{J_0}{r^{2\alpha}} G_x^{ab} G_{x+r}^{ab}]\\&- \frac{\delta(t_{12})}{4} [ V_0 (G_x^{ab})^2 + V_1 G_x^{ab}G_{x+1}^{ab}] \Big].
\end{aligned}
\end{equation}
It exhibits} $O(2)\times O(2)$ symmetry $G\rightarrow OGO^{T}$ and $\Sigma\rightarrow O\Sigma O^{T}$ with $O=\exp(-\gamma_{13}\theta_{13}-\gamma_{24}\theta_{24})$ and $(\gamma_{cd})^{ab}=\delta_{ac}\delta_{bd}-\delta_{bc}\delta_{ad}$. This owes to the enlarge of the permutation symmetry (between $1,3$ or $2,4$) for quadratic actions \cite{part_one}. In the large-$N$ limit, the saddle point solution for Green's function $G^{ab}_s$ is
\begin{equation}\label{eq:GBrownian}
\begin{aligned}
&G_s^{11}=-G_s^{22}=\frac{i\omega}{\omega^2+\Gamma^2/4},\\&G_s^{21}=-G_s^{12}=\frac{\Gamma/2}{\omega^2+\Gamma^2/4}.
\end{aligned}
\end{equation}
We also have $G_s^{a,b}=G_s^{a-2,b-2}$ for $a,b=3,4$, while other components of $G_s$ are zero. Here $\Gamma\equiv V+J$, with $V=V_0+V_1$ and $J=J_0(1+\sum_{r=1}^{\infty}1/r^{2\alpha})$. When $\alpha<1/2$, the summation diverges and we need to scale $J_0$ with system size $L$ to obtained a meaningful thermodynamic limit. This indicates the system becomes all-to-all connected, similar to the single SYK$_2$ model with $NL$ Majorana modes, and the steady state has volume-law entanglement regardless of the strength of the $V_0$ \cite{liu2018quantum,10.21468/SciPostPhys.8.6.094}. For $\alpha>0.5$, we are able to perform the summation as $J=J_0(1+\zeta(2\alpha))$, where $\zeta(x)$ is the Riemann zeta function. We will focus on this case in the following discussions.

\section{Effective Action}
 The saddle-point solution $G_s$ breaks the $O(2)\times O(2)$ symmetry down to $O(2)$. Consequently, the fluctuation around the saddle-point contains a gapless Goldstone mode \cite{part_one,part_two}. {We consider the fluctuation around saddle-point
\begin{equation}
\begin{aligned}
	\Sigma(t_1, t_2) &= \Sigma_s(t_1, t_2) + \delta \Sigma(t_1) \delta(t_{12}),\\ G(t_1, t_2) &= G_s(t_1, t_2) + \delta G(t_1, t_2),
	\end{aligned}
\end{equation}
and keep to the quadratic order of $\delta G$ and $\delta \Sigma$. We will focus on components between two replicas as in \cite{part_one}. Motivated by the symmetry analysis, we define $\theta(t)=\delta G^{14}(t,t)/2$ in the coset space. After integrating out gapped modes, the effective action derived in Appendix \ref{app:Brownianeff} reads }
\begin{equation}
\label{eq:effective_action_Brownian_theta}
\frac{I_{\text{eff}}}{N}=\frac{1}{2}\int \frac{d\omega dk}{(2\pi)^2}\left(\frac{\epsilon(k)}{2}+\frac{\omega^2}{4V}\right)|\theta(\omega,k)|^2,
\end{equation}
where the dispersion $\epsilon(k)$ is defined as
\begin{equation}\label{eqn_dispersion}
\epsilon(k)=V_1(1-\cos(k))+\sum_{r=1}^{\infty}\frac{J_0}{r^{2\alpha}}(1-\cos(rk)).
\end{equation}
Here we have set lattice constant $a_{\text{l}}=1$. To determine the low-energy effective theory, we expand $\epsilon(k)$ at small momentum $k\ll 1$. Two possibilities exist depending on $\alpha$:
\begin{equation}\label{eqn_epsilon}
  \epsilon(k) =
    \begin{cases}
      C_1 k^2/2 & \text{for $\alpha>1.5$,}\\
      C_2 k^{2\alpha-1} & \text{for $\alpha\leq1.5$.}
    \end{cases}       
\end{equation}
Here we have $C_1={V_1+J_0 \zeta(2\alpha-2)}$ and $C_2=-J_0\cos \left(\frac{1}{2} \pi  (1-2 \alpha )\right) \Gamma (1-2 \alpha )$. This can be understood by expanding $(1-\cos(rk))\approx k^2r^2/2$ and realizing the summation is convergent only for $\alpha>1.5$. { The presence of $k^{2\alpha-1}$ for $\alpha< 1.5$ leads to an anoumalous action \cite{lepori2016effective}, which breaks the conformal symmetry in the replicated Hilbert space.} The asymptotic behavior of $\epsilon(k)$ directly determines the dynamical exponent $z$. We have $z=1$ for $\alpha>1.5$ and $z=\frac{2\alpha-1}{2}$ for $\alpha\leq 1.5$.
	\begin{figure}[tb]
		\centering
		\includegraphics[width=1\linewidth]{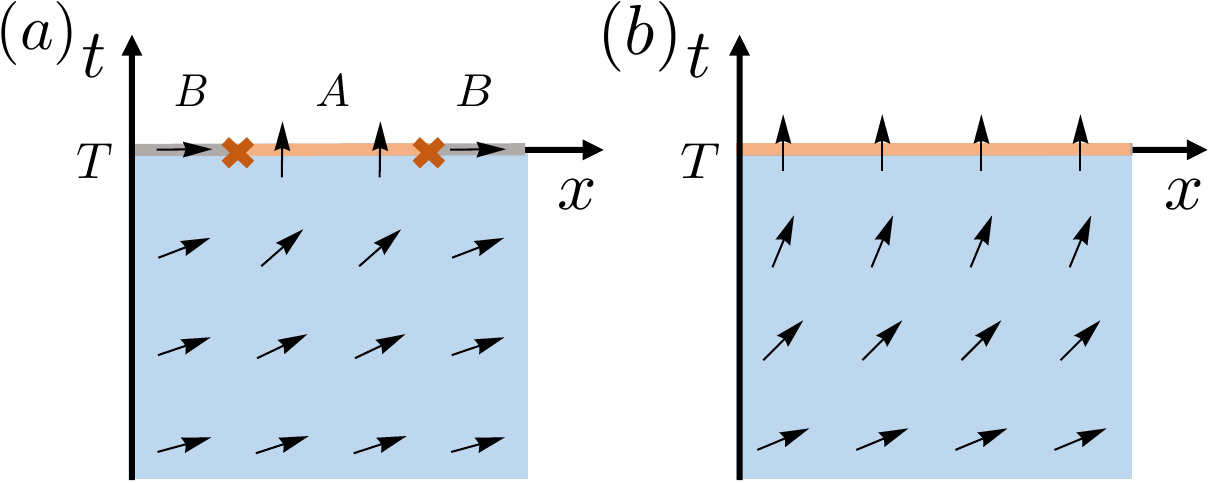}
		\caption{A sketch for the spin configuration ($\theta$ is the angle respect to the x-axis) that contributes to the calculation of (a). the second R\'enyi entanglement entropy on the steady state, (b). the purification dynamics. The cross represents the insertion of the twist operators.}
		\label{fig:entropy}
	\end{figure}

Now we use the effective action \eqref{eq:effective_action_Brownian_theta} to compute the Keldysh squared correlator $F$ and entanglement entropy $S^{(2)}_A$. On the replicated contour, $F$ corresponds to a four-point correlator of fermions, and thus a two-point function of collective fields $\theta$. More precisely, we have
\begin{equation}\label{eqn_Fscaling}
\begin{aligned}
F&\sim N\left<\partial_t\theta(x)\partial_t\theta(0)\right>_{11}\\&\sim\int_{\omega k} e^{ikx}\frac{\omega^2}{\omega^2+k^{2z}}\sim x^{-z-1}.
\end{aligned}
\end{equation}
Here we have dropped non-universal factors. 

For the entanglement entropy $S^{(2)}_A(T)$, the path-integral contour \eqref{eqn_purity} is defined with additional twist operators at $t=T$, as shown in FIG. \ref{fig:entropy} (a). In terms of $\theta$, this corresponds to fixing the boundary condition $\theta(T,x\in A)=\pi/2$ and $\theta(T,x\in B)=0$ \cite{part_one,part_two}, which creates a pair of half-vortices, separated by a distance $L_A$. $S^{(2)}_A$ is equal to the excitation energy of this half-vortex pair. This can be computed by the two-point correlator $P_A\sim\left<e^{i\pi \varphi(L_A)/2}e^{-i\pi \varphi(0)/2}\right>$. Here $\varphi(x)$ is defined by $\nabla\varphi=N\partial_t\theta/4V$, which shifts the value of $\theta(y)$ for $y<x$. The effective action of $\varphi$ can be determined by introducing a Lagrangian multipler $r$ to impose the relation $\nabla\varphi=\partial_t\theta$ and then integrate out $\theta$ and $r$.  Working out the details as in Appendix \ref{app:Brownianeff} gives
\begin{equation}\label{eqn_entropyscaling}
\begin{aligned}
S_A^{(2)}&\sim\left<\varphi(L_A)\varphi(0)\right>\\&\sim\int_{\omega k}~\frac{N}{k^{2}+k^{2+2z}/\omega^2}e^{ikL_A}\sim NL_A^{1-z}.
\end{aligned}
\end{equation}
For $z=1$, this should be understood as $\log L_A$. We find for $0.5<\alpha<1.5$, the system is subvolume-law entangled. This is called the fractal phase in \cite{ippoliti2021fractal}. It is straightforward to extend the above discussion to compute the mutual information between two small subregions separated by distance $d$. The result is $I^{(2)}(d)\sim \left<\nabla\varphi(d)\nabla\varphi(0)\right>\sim Nd^{-z-1}$, which matches the scaling of the squared correlator $F$.

It is also interesting to ask about the purification dynamics \cite{gullans2019dynamical} in the free fermion system and check how long it will take to purify the system \cite{Fidkowski_2021}. We prepare the system in the maximally mixed state and evolve it under the non-unitary dynamics for time $T$. The second R\'enyi entropy $S^{(2)}(T)$ of the full system then describes how much the initial quantum information stored in the system is lost \cite{gullans2019dynamical}. In the effective theory, $S^{(2)}(T)$ is given by computing the energy of the $\theta$ field with a boundary condition $\theta(t=0)=0$ and $\theta(t=T)=\pi/2$. The dominant contribution is determined by the saddle-point equation $\partial_t^2 \theta=0$, which gives $\theta(t)=\pi t/2T$. This configuration is shown in FIG. \ref{fig:entropy} (b), where the spins rotate smoothly from $0$ to $\pi/2$. This immediately suggests the purification process is insensitive to the range of the interaction. Explicitly, we have $S^{(2)}(T)/N\sim TL\times 1/T^2\sim L/T$ and the purification time $\sim O(L)$ for any arbitrary $\alpha\geq 0$.

\section{Single-flavor Model}
Now we ask whether our results derived in the large$-N$ limit hold for systems with small local Hilbert space dimension. To answer this question, we introduce a second model with single-flavor of fermion per site \cite{Complex}:
\begin{myequation}
H= \sum_{r\geq1} \left(\frac{J^{x,r+x}(t)}{r^\alpha} c^{\dagger}_{x+r}c_x+\text{H.C.}\right)-i\kappa_x(t)c^\dagger_xc_x.
\label{eq:single-flavor}
\end{myequation}
The first term describes the long-range hopping and the second term is an on-site imaginary potential. Similar to the SYK case, $J^{x,r+x}$ and $\kappa_x$ are Brownian variables with variance
\begin{equation}
\begin{aligned}
&\overline{J^{x,x+r}(t)J^{x,x+r}(0)^*}=J_l\delta(t),\\&\overline{\kappa_x(t)\kappa_x(0)}=\kappa\delta(t).
\end{aligned}
\end{equation}
We further choose $\kappa=1$ as the energy unit. At time $t$, 
\begin{align}
    |\psi(t)\rangle=\frac{e^{-iHt}}{\sqrt{\langle \psi_0|e^{iH^\dagger t}e^{-iHt}|\psi_0\rangle}}|\psi_0\rangle.
\end{align}
where the initial state $|\psi_0\rangle$ is chosen as a product state in the real space. It is known that under the non-unitary quadratic evolution, $|\psi(t)\rangle$ remains a fermionic Gaussian state \cite{bravyi2004lagrangian} and all information is encoded in the correlation function $C_{xy}(t)\equiv \left<\psi(t)\right|c^\dagger_xc_y|\psi(t)\rangle$. For the Brownian model, it is useful to introduce the distribution function as
\begin{align}
    f_n\equiv 
    \begin{cases}
    \sum_x \frac{|C_{x,x}|^2}{L}, &n=0\\
    \sum_{x}\frac{|C_{x,x+n}|^2+|C_{x,x-n}|^2}{L}, &n>0
    \end{cases}.
    \label{eq:fndef}
\end{align} 
As in \cite{Chen_2020}, $f_{n>1}$ approximately satisfies a set of non-linear master equation, which takes the form
\begin{equation}\label{eqn_masterequation}
\begin{aligned}
\frac{df_n}{dt} &=  \mu_n+\sum_{0<r<n}\frac{J_l}{r^{2\alpha}}(f_{n+r}+f_{n-r}-2f_n) \\&+\sum_{r\geq n}\frac{J_l}{r^{2\alpha}}(f_{n+r}-2f_n) 
- 2f_n \sum_{m=1}^\infty f_m  \\&+ \sum_{m=1}^\infty f_m f_{m+n} + \frac{1}{2}\sum_{m=1}^{n-1}f_m f_{n-m}.
\end{aligned}
\end{equation}
Here the first term $\mu_n= \mu_l/n^{2\alpha}$ is a source term which captures contributions from the diagonal term $C_{x,x}$. The second and the third term are contributions from the Hermitian Brownian hopping and describe a L\'evy flight process. Other terms are contributions from the random imaginary fields (see Appendix \ref{app:nonlinearmaster}) and is non-linear. For an initial product state, we have $f_{n>1}(0)=0$. 

We aim to understand the dynamics and the long-time behavior of \eqref{eqn_masterequation}. In the short-range hopping limit where $J^{x,x+r}=0$ for any $r\geq2$, it is known that the steady state has $f_n\sim 1/n^2$ and the dynamical exponent $z=1$ \cite{Chen_2020}. Now we turn on the long-range hopping term. Since the L\'evy flight is equivalent to a diffusive random walker when $\alpha>1.5$, we expect that the entanglement/correlation dynamics is the same as the system with only local hopping term. In this regime, the diffusive term is much slower than the non-linear terms in the master equation and can be safely removed when we analyze the dynamics of the master equation \cite{Chen_2020}.

	\begin{figure}[tb]
		\centering
		\includegraphics[width=1\linewidth]{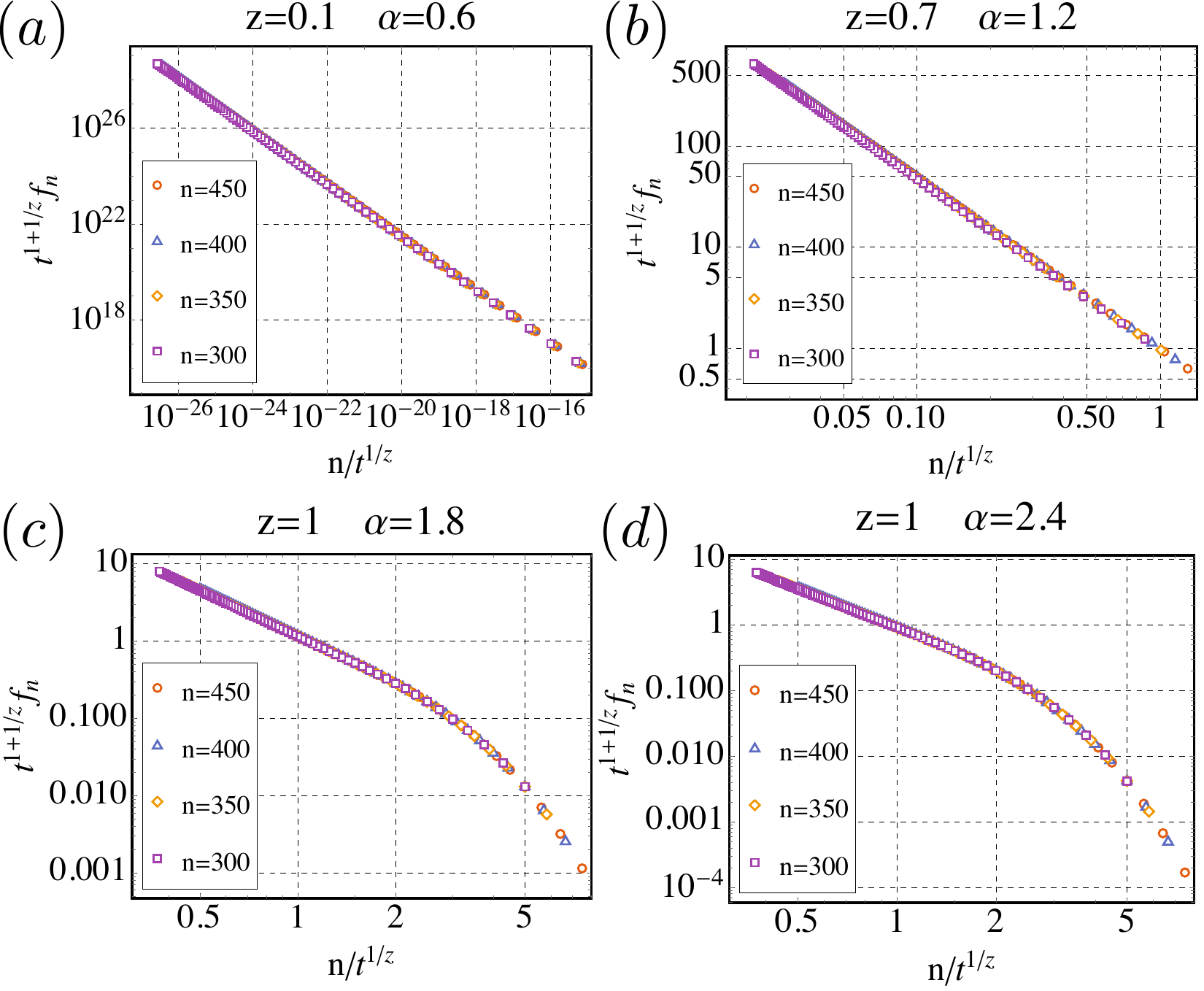}
		\caption{Numerical results obtained by solving the nonlinear master equation \eqref{eqn_masterequation} with $\mu_l=J_l=1/\zeta(2\alpha)$. We take $t\in [60,800]$ with a cutoff at $n=1000$. The slopes for the curves in (a) and (b) are -1.1 and -1.7 respectively. The results show the validity of the scaling form \eqref{scaling}.}
		\label{fig:numerics}
	\end{figure}

To determine the physics with $\alpha<1.5$, we first consider the steady state where $df_n/dt=0$. The R.H.S. of \eqref{eqn_masterequation} contains three terms. The contribution from $\mu_n$ is proportional to $1/n^{2\alpha}$. Assuming $f_n\sim 1/n^\delta$, the contribution from the imaginary potential gives
\begin{equation}\label{compare2}
\int dm~f_m(2\cos(m\partial_n)-2)f_n\sim  1/n^{2\delta-1}.
\end{equation}
The steady state is achieved when the contribution from the imaginary potential balances the contribution from $\mu_n$. This gives $\delta=\frac{2\alpha+1}{2}$. For any $\alpha>0.5$, this guarantees that the L\'evy flight terms have larger power-law exponents than the other two terms and can be safely neglected. Assuming the entanglement is contributed by EPR pairs with length distribution described by $f_n$ \cite{Nahum_2020}, we can perform a double integral over it to obtain the entanglement entropy $S^{(2)}_A\sim L_A^{{(3-2\alpha)/2}}$. All these results are consistent with the SYK calculation in the large-$N$ limit.

Now we turn to the determination of the dynamical exponent $z$ in the regime $0.5<\alpha<1.5$. At long but finite time $t$, we expect $f_n$ takes the scaling form:
\begin{equation}\label{scaling}
f_n(t)=t^{-\eta}F(n/t^{1/z}).
\end{equation}
Taking the infinite time limit, we should have 
\begin{equation}
f_n(t\to\infty)\sim t^{-\eta}(nt^{-1/z})^{-\eta z}\sim n^{-\eta z}.
\end{equation}
Consequently, we have the constraint $\eta z=\delta=\frac{2\alpha+1}{2}$. To determine $z$, we plug the scaling form \eqref{scaling} into equation \eqref{eqn_masterequation}. It is straightforward to confirm that $df_n/dt$ balances the non-linear convolution term from the imaginary potential \footnote{The last two terms in the master equation    \eqref{eqn_masterequation} combine into a convolution term in the continuum limit.}. We have $df_n/dt \sim 1/t^{\eta+1}$ and
\begin{equation}
\begin{aligned}
\int dm~f_m(2\cos(m\partial_n)-2)f_n\sim 1/t^{2\eta-1/z}.
\end{aligned}
\end{equation}
We conclude $2\eta-1/z=\eta+1$ and thus $z=\frac{2\alpha-1}{2}$. This again matches the large$-N$ result. To verify the scaling form \eqref{scaling}, we numerically solve the differential equation \eqref{eqn_masterequation}. The results are shown in Fig.~\ref{fig:numerics}, from which it is clear that we have $z=1$ for $\alpha>1.5$ and $z=\frac{2\alpha-1}{2}$ for $0.5<\alpha<1.5$, consistent with our analysis \footnote{We also numerically check the purification dynamics and find that $S(T)\sim L/T$, the same as the large-$N$ system.}. In particular, we find that the L\'evy flight terms can be disregarded (set $J_l=0$) and does not change the dynamics and the steady state, the same as the master equation with local diffusion term \cite{Chen_2020}. 

\section{Static Hopping Case} 
Now we examine whether our phase diagram works for time independent Hamiltonians. The large$-N$ model \eqref{eqn_HSYK} can be made static by assuming that the random variables are time-independent:
\begin{equation}
\begin{aligned}
&\overline{(\tilde{J}_{ij}^{x})^2}=\frac{J_0^2}{N},\ \ \ \ \ \ \ \overline{(V_{ij}^{x,x+1})^2}=\frac{V_1^2}{2N},\\&\overline{(\tilde{V}_{ij}^{x})^2}=\frac{V_0^2}{N},\ \ \ \ \ \ \ \overline{(J_{ij}^{x,x+r})^2}=\frac{J_0^2}{2N}.
\end{aligned}
\end{equation}
This model can also be analyzed analogously to the Brownian case (see Appendix \ref{app:staticmodel}). The main difference is that now the effective action contains copies of Goldstone modes because of the enlarged symmetry due to the time translation symmetry \cite{part_one}. For each copy, its effective action takes the same form as \eqref{eq:effective_action_Brownian_theta}. Consequently, the scaling of the $F$ in \eqref{eqn_Fscaling}, the scaling of $S^{(2)}_A$ in \eqref{eqn_entropyscaling}, and thus phase diagram in FIG. \ref{fig:schemticas} are still valid. We also numerically study the static version of $N=1$ single-flavor model in \eqref{eq:single-flavor}. Here we take hermitian long-range hopping terms to be time independent while keeping the imaginary potential to be random in the time direction. We manage to numerically reproduce the same result for $\alpha>1.5$ in Appendix \ref{app:numerics}. Putting all results together, we conclude our phase diagram FIG. \ref{fig:schemticas} is universal for both Brownian and static models, regardless of the local Hilbert space dimension. This is consistent with  recent studies \cite{minato2021fate,noteadded}.

\section{Discussions}
 In this work, we consider the long-range non-unitary random dynamics in free-fermion systems. We analytically show that both the large-$N$ Brownian/static SYK$_2$ chain and the single-flavor Brownian model exhibit a logarithmic phase for $\alpha>1.5$ where entanglement is logarithmic in subsystem size, and a fractal phase with $0.5<\alpha<1.5$ where the entanglement is subvolume-law. We also show that these two phases cannot be distinguished by the purification dynamics. We
 expect that the phase diagram is universal for general non-interacting random fermionic systems under long-range non-unitary dynamics. {The non-unitary evolution can be interpreted as the  dynamics under continuous forced measurements, as explained in Appendix \ref{app:exp}. An explicit experimental scheme with continuous measurement of particle density on each site has been proposed in \cite{hurst2019measurement} using interactions between atoms and photons. Consequently, both entanglement entropy \cite{islam2015measuring} and squared correlator can, in principle, be measured in an experiment with certain post-selection.}

 We finally make a few comments. Firstly, it is interesting to extend the discussions to general spatial dimension $D$. For the SYK$_2$ model, this corresponds to changing the summation over $r$ in \eqref{eqn_dispersion} to a $D$-dimensional integral. Consequently, for $\alpha>\frac{2+D}{2}$ we have a phase with $z=1$, $F(x)=1/x^{D+1}$ and $S^{(2)}=L_A^{D-1}\log L_A$, while for $\frac{2+D}{2}>\alpha>\frac{D}{2}$, we have $z=\frac{2\alpha-D}{2}$, $F(x)=1/x^{D+z}$ and $S^{(2)}=L_A^{D-z}$. Secondly, in \cite{part_two}, authors study the measurement effect on SYK chains by introducing models with two copy of chains with interchain coupling $\mu$. We can construct a similar model with long-range hopping terms. When $\alpha>0.5$, a transition between the fractal/logarithmic phase and the area law phase occurs at $\mu=J_0(1+\zeta(2\alpha))$, similar to the observed phase diagram in \cite{minato2021fate}.

\textit{Acknowledgment.} PZ acknowledges support from the Walter Burke Institute for Theoretical Physics at Caltech. SKJ is supported by the Simons Foundation via the It From Qubit Collaboration. CL is supported by the NSF CMMT program under Grants No. DMR-1818533. 

\textit{Note added.} After this work had been completed, we became aware of an independent investigation of entanglement transition in long-range hopping fermion chains~\cite{noteadded}.

\bibliographystyle{unsrtnat}
\bibliography{quantumver.bbl}

\onecolumn\newpage
\appendix

\section{Effective action and Entanglement entropy for the Brownian SYK chain}\label{app:Brownianeff}
As we discussed in the main text, we consider the replicated system $\overline{(\left<\psi_0\right| e^{iH^\dagger t} e^{-iHt}\left|\psi_0\right>)^2}$. After introducing bilocal fields and integrating out Majorana fermions, the $G$-$\Sigma$ action reads
\begin{equation} \label{sm:full_action}
\begin{aligned}
- \frac{I}N =& \sum_x \frac12 \text{tr} \log \Big( (-1)^{a+1} \delta^{ab} \partial_t - \Sigma_x^{ab} \Big)  - \int_{tt'} \Big[ \frac12 \Sigma_x^{ab} G_x^{ab} +   \frac{(-1)^{a + b}\delta(t-t')}4 [J_0 (G_x^{ab})^2 \\
&+\sum_r\frac{J_0}{r^{2\alpha}} G_x^{ab} G_{x+r}^{ab}]- \frac{\delta(t-t')}{4} [ V_0 (G_x^{ab})^2 + V_1 G_x^{ab}G_{x+1}^{ab}] \Big].
\end{aligned}
\end{equation}
We first consider the saddle-point equation for $G_x$ and $\Sigma_x$, which reads
\begin{equation}
\begin{aligned}
&\left[(-1)^{a+1}\delta^{ac}\partial_t-\Sigma_x^{ac}\right]\circ G_x^{cb}=I^{ab}.\\
&\Sigma_x^{ab}(t,t')=V_1\delta(t-t')\frac{G_{x+1}^{ab}(t,t')+G_{x-1}^{ab}(t,t')}{2}-\sum_{r\geq 1}(-1)^{a+b}\frac{J_0}{r^{2\alpha}}\frac{G_{x+r}^{ab}(t,t')+G_{x-r}^{ab}(t,t')}{2}\\&\ \ \ \ \ \ \ \ \ \ \ \ \ +V_0G_{x}^{ab}(t,t')-J_0(-1)^{a+b}G_{x}^{ab}(t,t').
\end{aligned}
\end{equation} 
The solution of the equation is translational invariant along space or time $\Sigma_x^{ab}(t,t')=G_s^{ab}(t-t')$. In fact, it takes exactly the same form as the model without long-range hopping \cite{part_one}:
\begin{equation}\label{eq:GBrownian2}
G_s^{11}(\omega)=\frac{i\omega}{\omega^2+\Gamma^2/4},\ \ \ \ G_s^{12}(\omega)=-\frac{\Gamma/2}{\omega^2+\Gamma^2/4},
\end{equation}
We also have $G_s^{a,b}=G_s^{a-2,b-2}$ for $a,b=3,4$, while other components of $G_s$ are zero. The effective action can be derived by consider fluctuations around the saddle-point,
\begin{equation}
	\Sigma(t_1, t_2) = \Sigma_s(t_1, t_2) + \delta \Sigma(t_1) \delta(t_{12}), \quad G(t_1, t_2) = G_s(t_1, t_2) + \delta G(t_1, t_2),
\end{equation}
and keep everything to the quadratic order. We will focus on components between two replicas. There are contributions from the $\text{tr} \log$ term, the $\Sigma G$ term, and the $G^2$ terms. Comparing to the short-range hopping model \cite{part_one}, the only difference is from the contribution from $G^2$ terms. Fortunately, these contributions are the simplest ones. As in \cite{part_one}, we introduce
\begin{equation}
\begin{aligned}
	\phi_1(\omega) &= \sqrt{2}\int dt \delta G^{13}(t,t) e^{i \omega t}, \\
	\phi_2(\omega) &= -\sqrt{2}\int dt \delta G^{14}(t,t) e^{i \omega t}.
	\end{aligned}
\end{equation}
The $G^2$ term gives a contribution
\begin{equation}
-\frac{\delta I_{\text{eff}}}N=\frac{1}{2}\int_t \left( V_k(\phi_{1,k} \phi_{1,-k} + \phi_{2,k}\phi_{2,-k})  - J_k (\phi_{1,k} \phi_{1,-k}- \phi_{2,k}\phi_{2,-k}) \right).
\end{equation}
Here $V_k=V_0+V_1 \cos(k)$, $J_k=J_0(1+\sum_r\cos(rk)/r^{2\alpha})$. Adding back contributions from other terms, we have
\begin{equation}
\label{sm:effective}
		-\frac{I_{eff}}N = \frac{1}{2}\sum_k \int_\omega \phi_k(\omega) \left( \ba{cccc} \Gamma-J_k  + V_k & -i \omega \\ i\omega & -\Gamma + J_k  + V_k\ea \right) \phi_{-k}(-\omega).
\end{equation}
Denoting $\theta=\sqrt{2}\phi_2$ and integrating out $\phi_1$, we arrive the effective action used in the main text:
\begin{equation}
\label{sm:effective_action_Brownian_theta}
\frac{I_{\text{eff}}}{N}=\frac{1}{2}\int \frac{d\omega dk}{(2\pi)^2}\left(\frac{\epsilon(k)}{2}+\frac{1}{4V}\omega^2\right)|\theta(\omega,k)|^2.
\end{equation}

The squared correlator $F$ corresponds to the two point function of $\phi_1$ \cite{part_one}. In the low energy limit, the equation of motion of $\phi_1$ gives $2V\phi_1(\omega)\approx -i\omega \phi_2(\omega)$. This justifies the calculation of $F$ in the main text.

For computing the entanglement entropy, as explained in the main text, we introduce Lagrangian multiplier $r(\omega,k)$ to impose the relation between $\theta$ and its dual field $\varphi$:
 \begin{equation}
\label{new}
I_{\text{eff}}=\int_{\omega k}\frac{N}{2}\left(k^{2z}+\omega^2\right)|\theta(\omega,k)|^2+r(\omega,k)(k\theta(\omega,k)+N\omega \varphi(\omega,k)).
\end{equation}
Here we have dropped non-universal parameters. Integrating out $r$ and $\theta$ leads to the effective action of $\varphi$ used the in main text:
\begin{equation}
\frac{I_{\text{eff}}}{N}=\frac{1}{2N}\int_{\omega k}\left(k^{2}+k^{2+2z}/\omega^2\right)|\varphi(\omega,k)|^2.
\end{equation}

  \section{Non-linear master equation for the single-flavor model} \label{app:nonlinearmaster}
Now we present the derivation of the non-linear master equation for the single-flavor model. The derivation also follows the short-range hopping case studied in \cite{chen2020emergent}.

We first write out the equation satisfied by the correlation matrix $C_{xy}$. Since the model is Brownian, we can obtain the evolution of $|C_{xy}|^2$ by using the It\^{o} calculus. The derivation is tedious but straightforward  \cite{chen2020emergent}. Here we only cite the result. The contribution from the
Hermitian long-range hopping part reads
\begin{equation}
\begin{aligned}
\left(\frac{d|C_{x,y}|^2}{dt}\right)^{(1)}=&\sum_r\frac{J_l}{r^{2\alpha}}\Big\{|C_{y+r,x}|^2+|C_{y-r,x}|^2+|C_{y,x+r}|^2+|C_{y,x-r}|^2\\
& -2\delta_{x,y+r}C_{y+r,y+r}C_{y,y}-2\delta_{x,y-r}C_{y-r,y-r}C_{y,y}\\
&-4|C_{x,y}|^2+2(C_{x+r,x+r}C_{x,x}+C_{x-r,x-r}C_{x,x})\delta_{x,y}\Big\}.
\label{sm:unitary_eom}
\end{aligned}
\end{equation}
The contribution from the on-site non-Hermiticity reads
\begin{equation}
\begin{aligned}
\left(\frac{d|C_{x,y}|^2}{dt}\right)^{(2)}=&4|C_{x,y}|^2+4|C_{x,y}|^2\delta_{a,b}-8|C_{x,y}|^2(C_{y,y}+C_{x,x})+4\sum_m|C_{y,m}|^2|C_{m,x}|^2\\
&+4\sum_m \left[C_{y,m}C_{m,x}C_{m,m}C_{x,y}+ C_{x,m}C_{m,y}C_{m,m}C_{y,x}\right]\\&-2\sum_m\left[ C_{y,m}C_{m,x}C_{x,y}+ C_{x,m}C_{m,y}C_{y,x}\right].
\label{sm:imaginary_eom}
\end{aligned}
\end{equation}
To study the dynamics of $f_n$, we rewrite Eq.\eqref{sm:unitary_eom} and Eq.\eqref{sm:imaginary_eom} in terms of $f_n$. Note that this is only possible under certain approximations. The Hermitian part Eq.\eqref{sm:unitary_eom} contributes a term 
$\sum J_l(f_{n-r}+f_{n+r}-2f_n)/r^{2\alpha}$. In particular, for $n=r$ this leads to a production of long-range correlation directly from the on-site correlation.
For Eq.\eqref{sm:imaginary_eom}, we follow the discussion in \cite{chen2020emergent} and throw away terms $C_{y,m}C_{m,x}C_{x,y}$ and $C_{y,m}C_{m,x}C_{m,m}C_{x,y}$ when $m\neq x$ or $y$. The contribution from the on-site imaginary potential then mainly contributes a convolution term $- 2f_n \sum_{m=1}^\infty f_m + \sum_{m=1}^\infty f_m f_{m+n} + \frac{1}{2}\sum_{m=1}^{n-1}f_m f_{n-m}$. Summing up these contributions, we arrive at the master equation used in the main text 
\begin{equation}\label{sm_masterequation}
\begin{aligned}
\frac{df_n}{dt} = & \mu_n+\sum_{0<r<n}\frac{J_l}{r^{2\alpha}}(f_{n+r}+f_{n-r}-2f_n) +\sum_{r\geq n}\frac{J_l}{r^{2\alpha}}(f_{n+r}-2f_n) 
\\&- 2f_n \sum_{m=1}^\infty f_m + \sum_{m=1}^\infty f_m f_{m+n} + \frac{1}{2}\sum_{m=1}^{n-1}f_m f_{n-m},
\end{aligned}
\end{equation}

  \section{The static model}\label{app:staticmodel}
We first consider the static SYK chain with long-range non-Hermitian Hamiltonian. The self-consistent equation for the Green's function reads
\begin{equation}
\begin{aligned}
&\left[(-1)^{a+1}\delta^{ac}\partial_t-\Sigma_x^{ac}\right]\circ G_x^{cb}=I^{ab}.\\
&\Sigma_x^{ab}=V_1^2\frac{G_{x+1}^{ab}+G_{x-1}^{ab}}{2}+V_0^2G_{x}^{ab}-(-1)^{a+b}J_0^2G_{x}^{ab}-\sum_{l\geq 1}(-1)^{a+b}\frac{J_0^2}{l^{2\alpha}}\frac{G_{x+l}^{ab}+G_{x-l}^{ab}}{2}.
\end{aligned}
\end{equation}
For the saddle-point solution with translational symmetry $G^{ab}_x=G_s^{ab}$. Focusing on the low-energy limit $|\omega|<\frac{2J^2}{\sqrt{J^2+V^2}}$, $G_s(\omega)$ reads
\begin{equation}\label{sm:GStatic}
G_s^{11}(\omega)=\frac{i\omega}{2J^2},\ \ \ \ G_s^{12}(\omega)=-\frac{1}{2J^2}\sqrt{\frac{4J^4}{J^2+V^2}-\omega^2},
\end{equation}
together with $G_s^{22}(t)=-G_s^{11}(t)$ and $G_s^{21}(t)=-G_s^{12}(t)$. Here we have defined $V^2\equiv V_0^2+V_1^2$ and
\begin{equation}
\begin{aligned}
J^2=J_0^2+\sum_{l\geq 1}\frac{J_0^2}{l^{2\alpha}}=J_0^2(1+\zeta(2\alpha)).
\end{aligned}
\end{equation} 
Now we turn to the effective action. The $G-\Sigma$ action reads
\begin{equation} \label{sm:staticfull_action}
\begin{aligned}
- \frac{I}N =& \sum_x \frac12 \text{Tr} \log \Big( (-1)^{a+1} \delta^{ab} \partial_t - \Sigma_x^{ab} \Big)  + \int \Big[ - \frac12 \Sigma_x^{ab} G_x^{ab}  \\
&+ \frac{1}{4} [ V_0^2 (G_x^{ab})^2 + V_1^2 G_x^{ab}G_{x+1}^{ab}]-   \frac{(-1)^{a + b}}4 [J_0^2 (G_x^{ab})^2 +\sum_l\frac{J_0^2}{l^{2\alpha}} G_x^{ab} G_{x+l}^{ab}] \Big],
\end{aligned}
\end{equation}
Expanding $G^{ab}_x(t,t')=G^{ab}_s(t,t')+\delta G^{ab}_x(t,t')$. We again focus on fluctuations involving two replicas $\{\delta G^{13}$, $\delta G^{14}$, $\delta G^{23}$, $\delta G^{24}\}$ and define $\phi_\pm =(\phi_\pm^1,\phi_\pm^2)= \frac1{\sqrt2}(\delta G^{13} \pm \delta G^{24}, \delta G^{14} \pm \delta G^{23}) $ as in \cite{part_one}, where it is found that only $\phi_+$ contributes in the low-energy limit. Keeping to the quadratic order as in the Brownian case, we find the effective action
\begin{equation}
\begin{aligned}
\label{sm:effective_action_static}
&-\frac{I_{\text{eff}}}{N}=\frac{1}{2}\int_{\Omega\omega k}\phi_+(\Omega,\omega,k)
\begin{pmatrix}
2V^2&\frac{i (J^2 + V^2)^{3/2}}{2J^2}\Omega\\
-\frac{i (J^2 + V^2)^{3/2}}{2J^2}\Omega&-\epsilon(k)+ \frac{(J^2+V^2)^2\Omega^2}{8J^4}
\end{pmatrix}
\phi_+(\Omega,\omega,k).
\end{aligned}
\end{equation}
Here $\Omega$ is the center-of-mass frequency and $\omega$ is the relative frequency of $t$ and $t'$. We have 
\begin{equation}
\epsilon(k)=\Big(1-\cos(k)\Big)V_1^2+\sum_{l\geq 1} \Big(1-\cos(lk)\Big)J_l^2\approx k^2+k^{2 \alpha
   -1},
\end{equation}
 which takes the same form as the Brownian case. Integrating out $\phi_+^1$ and using the coset variables \cite{part_one} leads to the anisotropic XY model:
\begin{equation}
\label{am:effective_action_static_theta}
\frac{I_{\text{eff}}}{N}=\frac{1}{2}\int_{\omega\Omega k}\left(\frac{2\epsilon (k)}{J^2+V^2}+\frac{J^2+V^2}{4J^2V^2}\Omega^2\right)|\theta^{\omega}(\Omega,k)|^2.
\end{equation}
This is the same for as the Brownian case, with an additional label $\omega$ for different Goldstone modes. Here we should treat $\Omega$ as the frequency in the Brownian case. Consequently, the scaling of the squared correlators and entanglement entropy should be the same as the Brownian model.

\section{Numerical simulation for $N=1$}\label{app:numerics}

For numerical convenience, we construct a {\it discrete} non-unitary circuit model for single-flavor free fermion with the evolution operator (unnormalized)
\begin{align}
U=\prod_{t=1}^{T} U_\beta(t)U_\tau(t).
\label{sm:non_unitary}
\end{align} 
Here $U_\tau=\exp(-iH_R\tau)$ represents the evolution governed by a long-range hopping Hermitian Hamiltonian $H_R$ and and $U_\beta=\exp(-H_I\beta)$ represents the imaginary evolution with Hamiltonian $H_I$. At time $t$, we have the wave function
\begin{align}
    |\psi(t)\rangle=\frac{U}{\sqrt{Z}}|\psi(0)\rangle,\quad \text{with}\ \ Z=\langle \psi(0)|U^\dag U|\psi(0)\rangle,
\end{align}
where the initial state $|\psi(0)\rangle$ is chosen as a product state in real space. If $H_R$ and $H_I$ are both quadratic Hamiltonians, $|\psi(t)\rangle$ remains a Gaussian state under time evolution. All the information is encoded in the two point correlation function matrix $C_{xy}=\langle\psi(t)|c^\dag_xc_y|\psi(t)\rangle$.

\begin{figure*}[t]
\centering
\subfigure[]{\label{fig:EE_alpha_18} \includegraphics[width=.4\columnwidth]{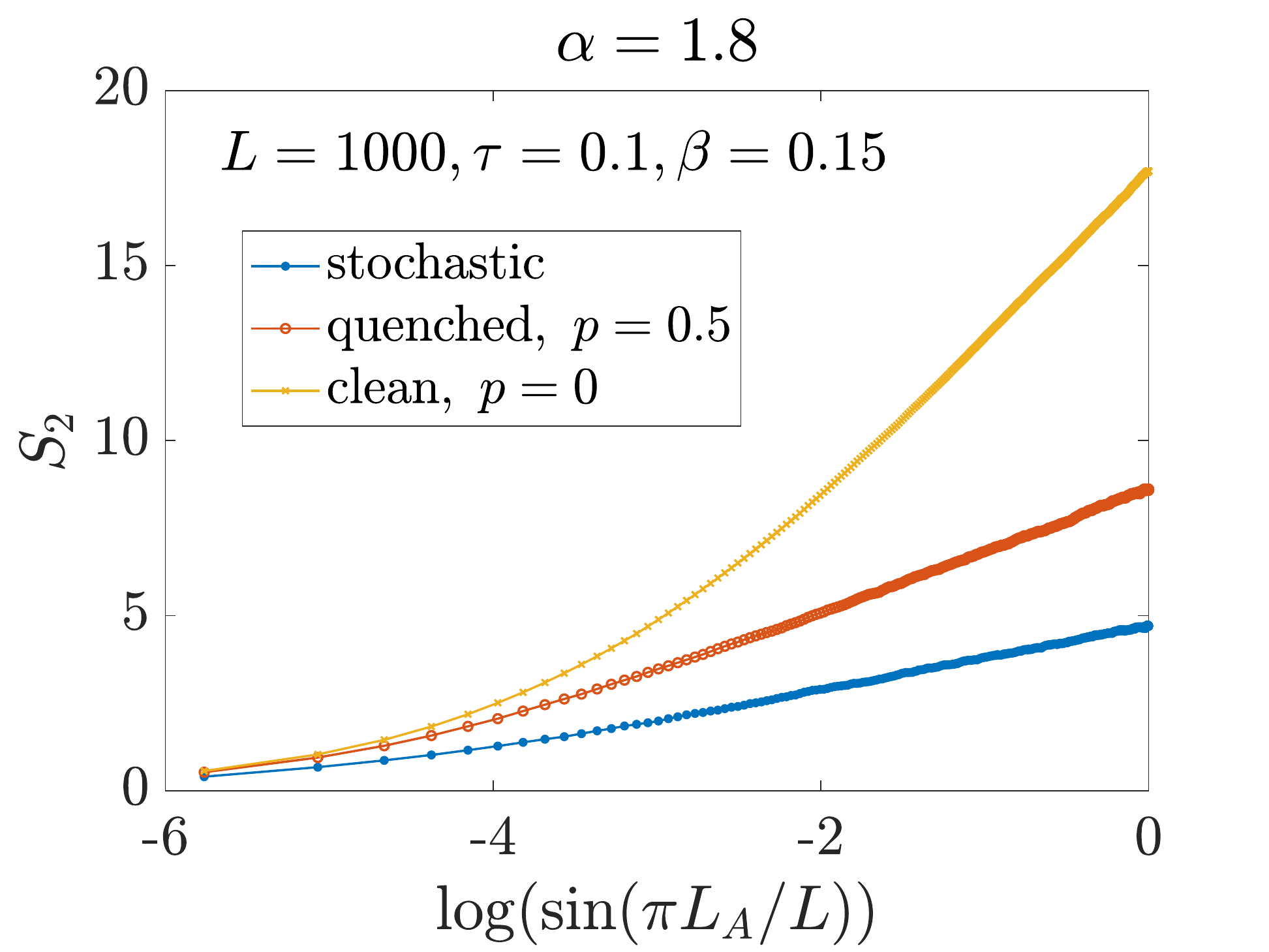}}
\subfigure[]{\label{fig:Corr_alpha_18} \includegraphics[width=.4\columnwidth]{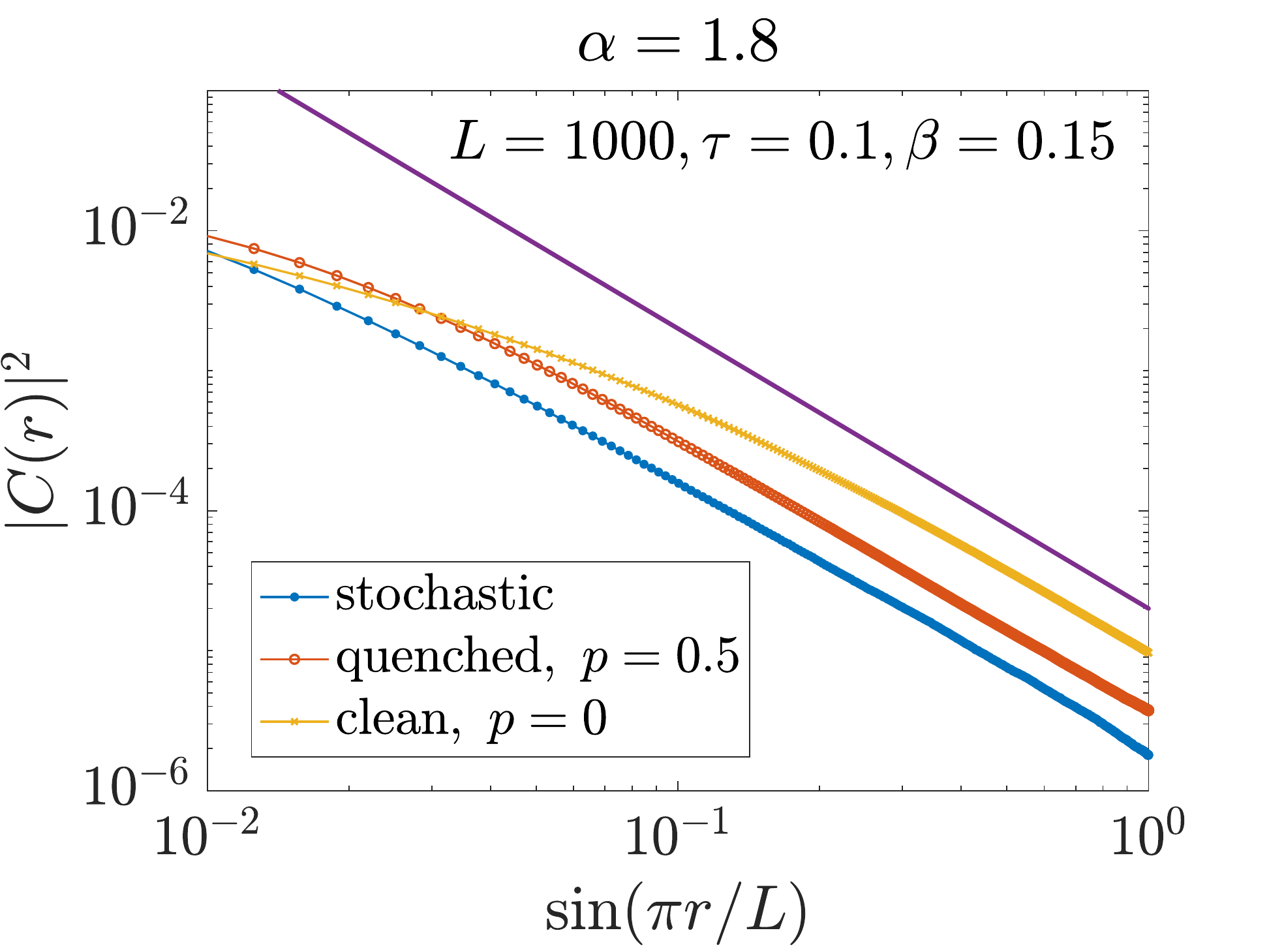}}
\subfigure[]{\label{fig:Dyn_alpha_18} \includegraphics[width=.4\columnwidth]{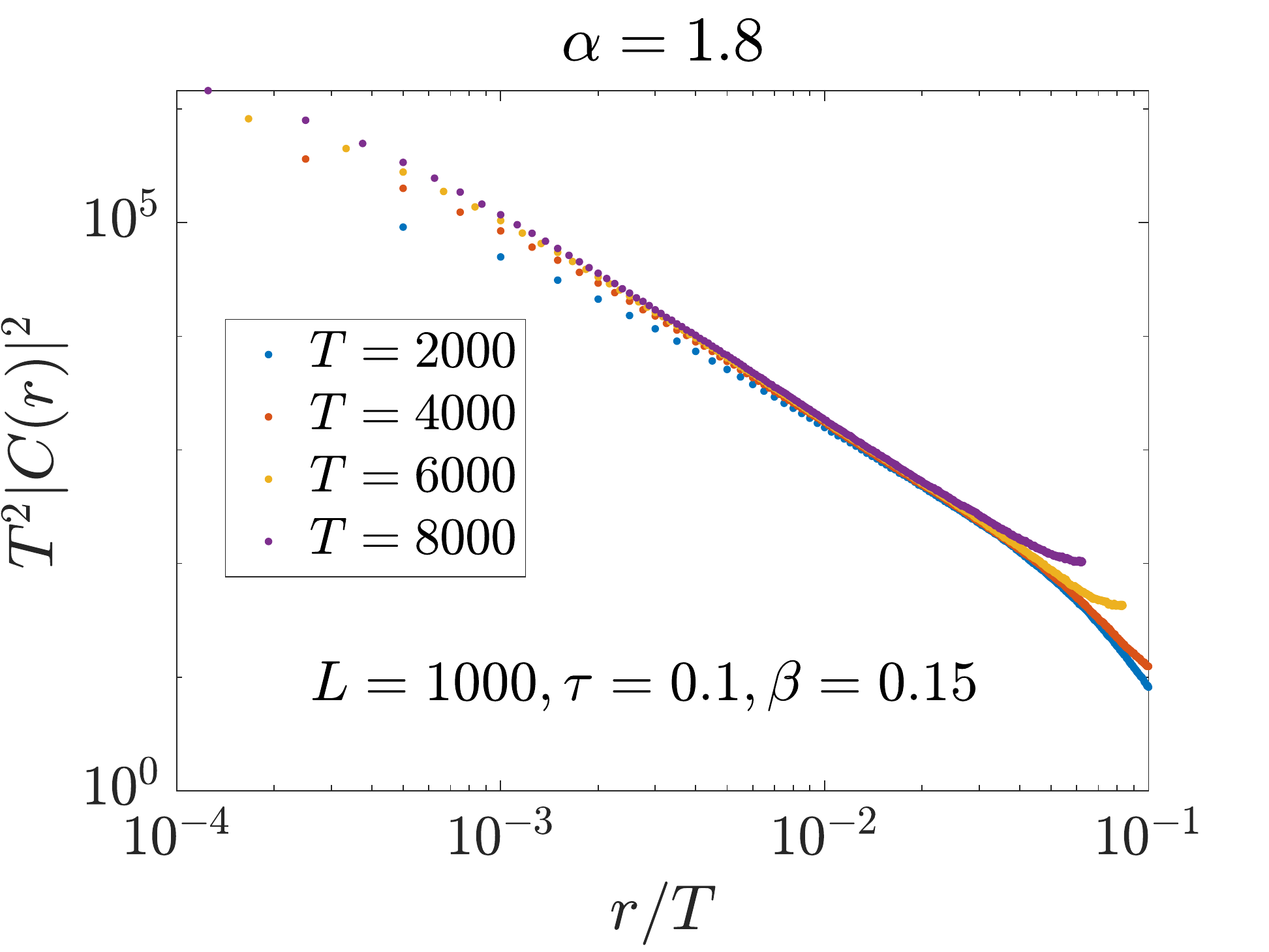}}
\subfigure[]{\label{fig:EE_com_alpha_18} \includegraphics[width=.4\columnwidth]{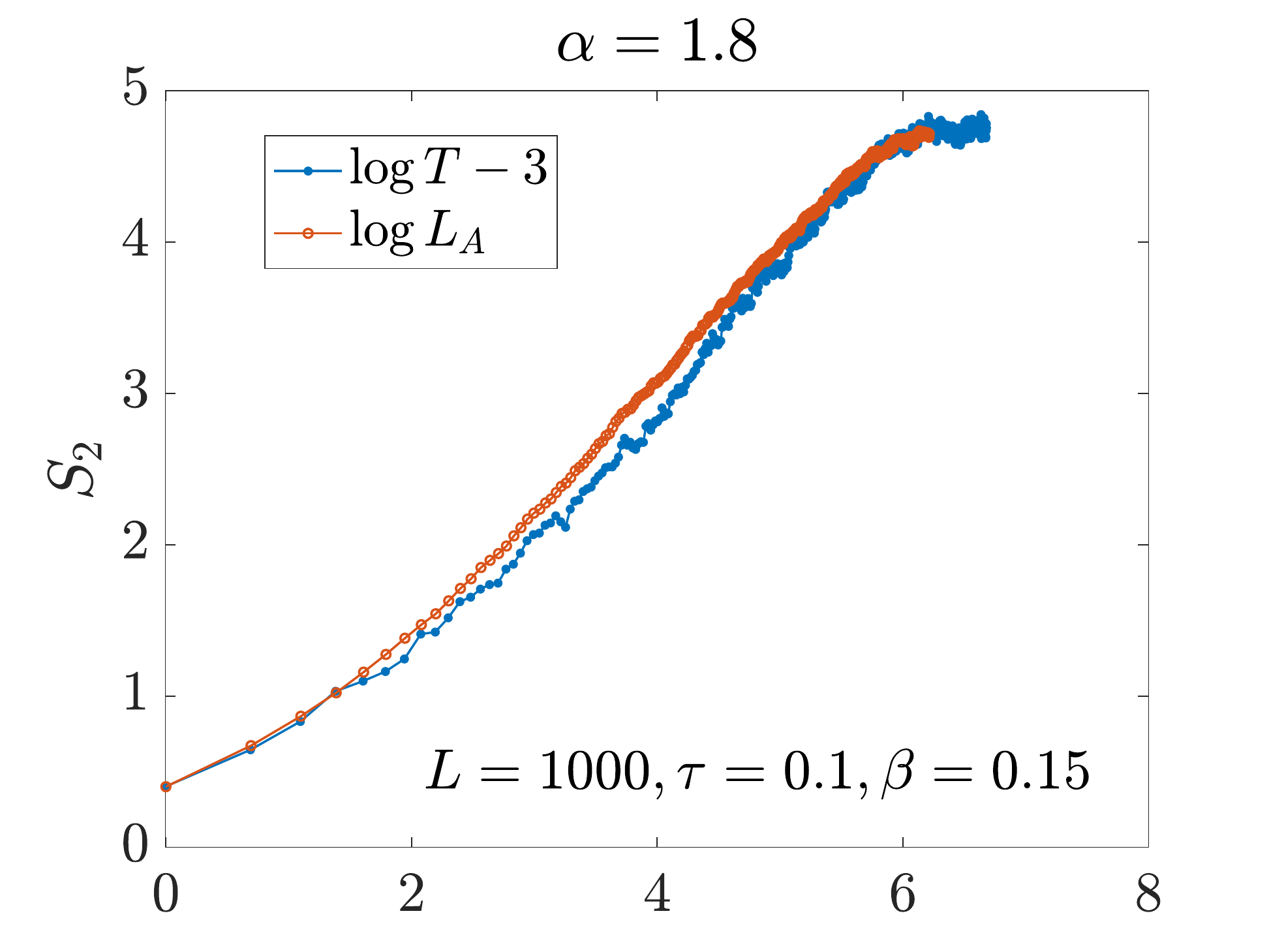}}
\caption{The numerical results at $\alpha=1.8$ with periodic boundary condition. All the results presented here are averaged over large number of samples. (a) The steady state second R\'enyi entropy for various non-unitary dynamics. For the blue curve, $J^{x,r+x}$ is a time dependent random variable with the distribution function described by \eqref{eq:stochastic_dis}. For the red curve, $J^{x,r+x}$ is a time independent random variable with the distribution function described by \eqref{eq:quenched_dis} with $p=0.5$. For the yellow curve, $J^{x,r+x}=1$ and $H_R$ has no randomness in it. (b) The steady state squared correlator for various non-unitary dynamics. The color of the curves are the same as that in panel (a). The purple curve is proportional to $1/\sin^2(\pi r/L)$. The blue, red and yellow curves are all parallel to the purple curve, suggesting that the squared correlator decays as $1/r^2$ in all these three models. (c) The data collapse of the squared correlator on the log-log scale. The curves at different times collapse into a single curve, suggesting that $z=1$. (d) The blue curve is the half system $S_2$ vs $\log T-3$ and the red curve is the steady state $S_2$ vs $\log L_A$. These two curves have the same slope, suggesting that $z=1$.} 
\label{fig:alpha_18}
\end{figure*}

To incorporate the long-range hopping and periodic boundary condition, we take the following Hamiltonian
\begin{equation}
\begin{aligned}
H_R&=\sum_{r\geq0}\frac{J^{x,r+x}}{1+\frac{\sin^{\alpha} (\pi r/L)}{ (\pi/L)^{\alpha}}} \left( c^{\dagger}_{x+r}c_x+\text{H.C.}\right)\\
H_I&=\kappa_x(t)c^\dagger_xc_x.
\end{aligned}
\end{equation}
Here $\kappa_x(t)$ is stochastic random variable with a simple two-component distribution,
\begin{align}
    P(\kappa_x(t))=\frac{1}{2}\delta(\kappa_x(t)-1)+\frac{1}{2}\delta(\kappa_x(t)+1).
\end{align}
For the variable $J^{x,r+x}$, we consider two versions: the stochastic and static. 

In the stochastic case, we take $J^{x,r+x}$ to be a time dependent random variable
\begin{align}
    P(J^{x,r+x}(t))=\frac{1}{2}\delta(J^{x,r+x}(t)-1)+\frac{1}{2}\delta(J^{x,r+x}(t)+1).
    \label{eq:stochastic_dis}
\end{align}
The numerical results for $\alpha=1.8$ is shown in Fig.~\ref{fig:alpha_18}. We show that this model has dynamical exponent $z=1$ and the final steady state has $S_2\sim \log L_A$ and $C(r)\sim 1/r^2$. We also take $H_R$ to be a time independent Hamiltonian with $J^{x,r+x}$ satisfying
\begin{align}
    P(J^{x,r+x})=p\delta(J^{x,r+x}-1)+(1-p)\delta(J^{x,r+x}+1).
    \label{eq:quenched_dis}
\end{align}
We present the steady state results for $p=0$ and $p=0.5$ in Fig.~\ref{fig:alpha_18} (a) (b). Notice that when $p=0$, $H_R$ is a clean system without disorder. They all have the same scaling behavior as the stochastic version. We also try other $\alpha>1.5$ and we obtain the same results. These numerical simulation results suggest that the discrete dynamics has the same result as the Brownian dynamics we discussed in the main text in the regime $\alpha>1.5$. However, we are unable to obtain consistent numerical results for $\alpha<1.5$ due to strong finite size effect with $L=1000$.

\subsection{Purification dynamics}
\begin{figure*}[t]
\centering
\includegraphics[width=.5\columnwidth]{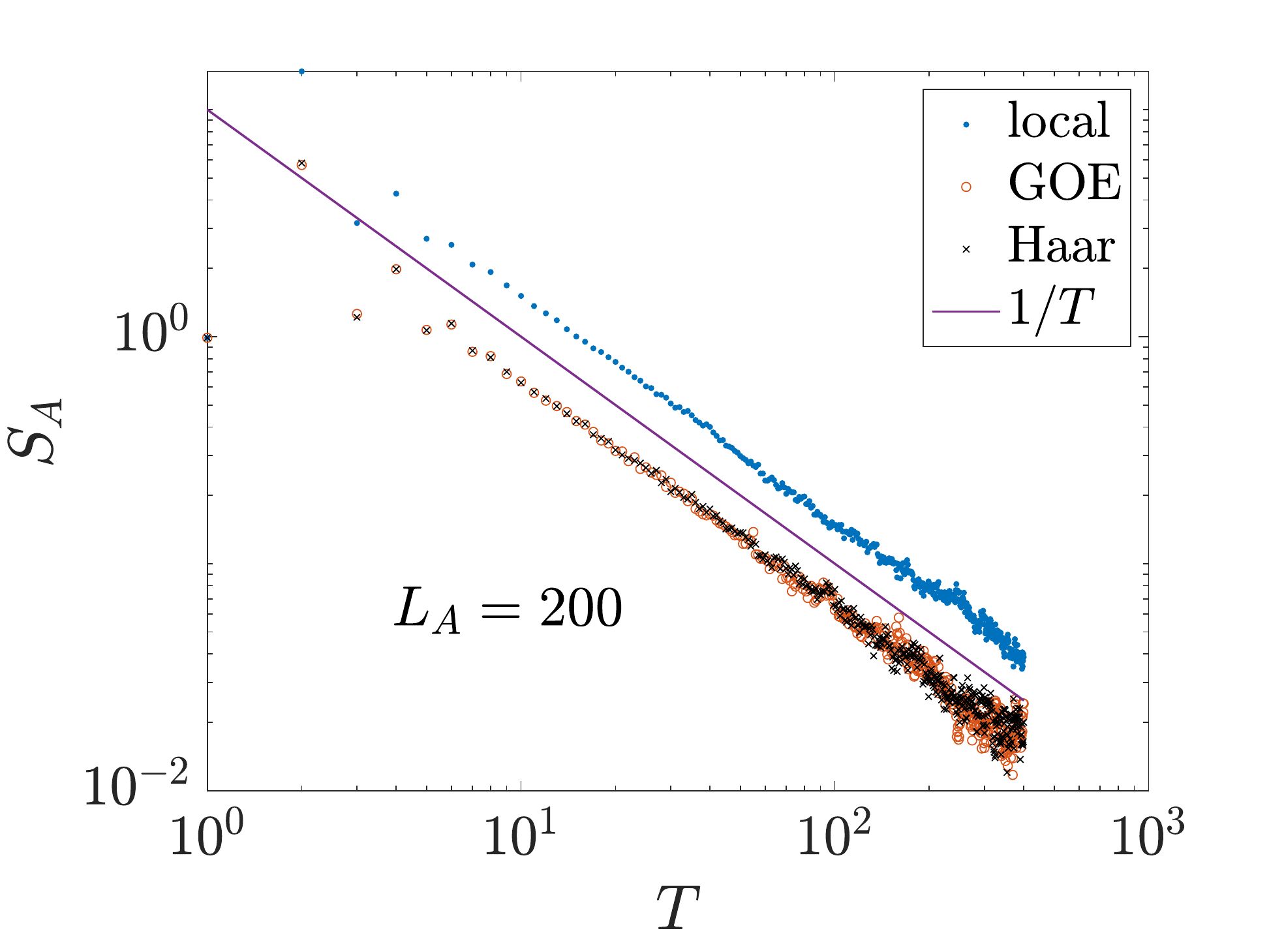}
\caption{The purification dynamics of non-unitary random dynamics. We take three different unitary evolutions $U_\tau$ and we show that the entropies of the system all decay as $1/T$.} 
\label{fig:purify}
\end{figure*}

We also take the initial system $A$ as a density matrix and study how fast it can purify under non-unitary random dynamics. We take the unitary dynamics $U_\tau$ to be random in the spatial and time direction and $H_I$ to be a stochastic random potential term. We consider three different cases for unitary dynamics : a local Hamiltonian for $H_R$, a single-particle Gaussian orthogonal ensemble (GOE) random matrix for $H_R$ and single-particle Haar random matrix for $U_\tau$. As shown in Fig.~\ref{fig:purify}, in all these cases, the system $A$ has the second R\'enyi entropy $S_A\sim 1/T$, regardless of the locality structure of $U_\tau$.

\section{The realization of non-unitary dynamics}\label{app:exp}
In this section, we discuss the experimental relevance of our work by relating the single quantum trajectory under the non-unitary evolution to evolution with continuously forced measurements. We focus on the single-flavor model where 
\begin{equation}\label{HI}
    H_I=\sum_x\kappa_x(t)c^\dagger_xc_x.
\end{equation}
After having generated a set of random numbers for $\kappa_x(t)$, we can make a constant shift of the potential such that $\kappa_x(t)>0$. This is due to the particle number conservation of the system. 

Intuitively, such imaginary potential terms should corresponds to measuring the on-site particle numbers. We hope to make such relation explicit. To begin with, we consider continuously monitoring our system: at each time step $dt$, we apply a measurement at site $x$. The measurement is described by a set of operators
\begin{equation}\label{RM}
    \{M_x^0,M_x^1\}=\left\{1-s_x(t)c^\dagger_xc_x,\sqrt{2s_x(t)-s_x(t)^2}c^\dagger_xc_x\right\}.
\end{equation}
Here we assume $0<s_x(t)\ll 1$, which means the measurement strength is weak. It is straightforward to check the completeness relation $\sum_iM_x^i(M_x^i)^\dagger=I$. For a system in some density matrix $\rho(t)$ before the measurement, after a single measurement, depending on the outcome, the unnormalized density matrix becomes
\begin{equation}
 \rho^0(t+dt)=M_x^0\rho(t)(M_x^0)^\dagger,\ \ \ \ \ \ \ \ 
    \rho^1(t+dt)=M_x^1\rho(t)(M_x^1)^\dagger.
\end{equation}
And the probability for obtaining $\rho^i(t+dt)$ is given by $p^i=\text{tr}[\rho^i(t+dt)]$. For forced measurement, we only keep the result when the outcome is $0$. Then the evolution of the unnormalized density matrix is given by
\begin{equation}
\begin{aligned}
   \rho(t+dt)~=~M_x^0\rho(t)M_x^0\approx  \rho(t)-s_x(t)c^\dagger_xc_x \rho(t)-\rho(t)s_x(t)c^\dagger_xc_x +o(s_x)
\end{aligned}
\end{equation}
Here we have kept only terms up to $O(s_x)$. Now we choose $s_x(t)=\kappa_x(t)dt$, the evolution can now be written as
\begin{equation}
\begin{aligned}
   \rho(t+dt)~=~e^{-\kappa_x(t)c^\dagger_xc_xdt} \rho(t)e^{-\kappa_x(t)c^\dagger_xc_xdt}+O(dt^2).
\end{aligned}
\end{equation}
After adding up measurements over all sites, this matches the result of non-Hermitian Hamiltonian dynamics. 

The traditional $n$-th subsystem R\'enyi entropy $S_A^{(n)}$ is defined as 
\begin{equation}\label{Sn}
    S_A^{(n)}= -\frac{1}{(n-1)}\text{tr}~\rho\log \left(\frac{\text{tr}_A(\text{tr}_B~\rho)^n}{(\text{tr}~ \rho)^n}\right)=-\frac{1}{(n-1)}\log \left(\frac{\text{tr}~[ \rho\otimes\rho...\otimes\rho\mathcal{X}_A]}{\text{tr}~[ \rho\otimes\rho...\otimes\rho]}\right).
\end{equation}
In particular, the second R\'enyi entropy $S^{(2)}_A$ has been measured in experiments \cite{islam2015measuring} for systems under unitary evolution. We have written the R\'enyi entropy as the expectation of the replicated system. Here $\mathcal{X}_A$ is the cyclic permutation operator for subsystem $A$.

We further show that $S_A^{(n)}$ can in principle be measured directly in experiments by generalization protocols in \cite{islam2015measuring}. For simplicity, we again focus on the $n=2$ case and analyze the effect of a single measurement. We prepare two copies of systems in some initial state $\rho(t)\otimes\rho(t)$. We now measure the system using operators $M_x^i\otimes M_x^j$: 
\begin{equation}\label{RM2}
    \{M_x^0\otimes M_x^0,M_x^1\otimes M_x^0,M_x^0\otimes M_x^1,M_x^1\otimes M_x^1\}.
\end{equation}
The probability of getting $ij$ is given by $\text{tr}~[\rho^i\otimes\rho^j]$. We now post select results with $i=j=0$ only. The probability of getting result $ii$ and the normalized state after getting the result are then given by
\begin{equation}
 \rho_d=\frac{\rho\otimes\rho}{\text{tr}~[\rho\otimes\rho]}.
\end{equation}
The swap operator $\mathcal{X}_A$ can be measured by applying additional Rabi oscillations between replicas \cite{islam2015measuring}. This leads to 
\begin{equation}
    \left<\mathcal{X}_A\right>=\text{tr}~[\mathcal{X}_A \rho_d]=\frac{\text{tr}~[ \rho\otimes\rho\mathcal{X}_A]}{\text{tr}~[ \rho\otimes\rho]},
\end{equation}
which directly gives $e^{-\tilde{S}_A^{(2)}}$. Note that this experimental realization faces the same challenging in the mixed unitary-measurement dynamics described in Refs. \cite{Cao_Tilloy_2019,Li_2018,Li_2019,Skinner_2019,Chan_2019,Bao_2020,Choi_2020,gullans2019dynamical,gullans2019scalable,jian2019measurementinduced,zabalo2019critical,Tang_Zhu_2020,Szyniszewski_2019,Zhang_2020,goto2020measurementinduced,jian2021yang,buchhold2021effective,bao2021symmetry}.
\end{document}